\documentclass{article}
\usepackage[english]{babel}
\usepackage[utf8]{inputenc}
\usepackage[T1]{fontenc}
\usepackage{graphicx}
\usepackage{longtable}
\usepackage{natbib}
\usepackage{longtable}
\usepackage[margin=3cm]{geometry}

\begin{document}

\title{\textbf{Light Curve for Type Ia Supernova Candidate AT 2018we}}

\author{T. Willamo$^{[1]}$, J. Ala-Könni$^{[1]}$, J. Arvo$^{[2]}$, I. Pippa$^{[1]}$, T. Salo$^{[1]}$}

\maketitle

\bigskip

[1] {Department of Physics, University of Helsinki, Finland}

[2] {Department of Philosophy, History and Art Studies, University of Helsinki, Finland}

\bigskip

email: teemu.willamo@helsinki.fi

\vspace{1cm}

\bigskip

\begin{abstract}

\noindent We present photometric follow-up observations of the supernova candidate AT 2018we, which was discovered by the Gaia photometric alerts system. The first Gaia detection of the object was made Feb 12, 2018. Our observations, made in Cousins R and Johnson B band, range from Feb 20 to May 5, 2018. From our observations, we have constructed a light curve, which supports the initial prediction of type Ia, included in the Gaia alert. Based on this classification, we have also estimated the distance to its assumed host galaxy, which was found to be around 43 Mpc.

\end{abstract}

\bigskip


\section{Introduction}

AT 2018we, also known as Gaia18alj, is a candidate supernova discovered by the Gaia Photometric Alert system near the galaxy shaped object GALEXASC J170959.67+820003.4 on Feb 12, 2018, at coordinates $\alpha$=17h9m57s, $\delta$=82$^\circ$0'3'' (J2000) \citep{delgado2018}; for the photometric alert page, see http://gsaweb.ast.cam.ac.uk/alerts/alert/Gaia18alj/. The comment "candidate SN near galaxy shaped object in DSS and PanSTARRS GS-TEC predicts SN Ia" is included in the alert. The GS-TEC prediction is made with the algorithms described in \cite{gs-tec}. They state that the 'purity' of Ia predictions of magnitude 19 (the magnitude of AT 2018we when the GS-TEC prediction was made) is 99.3\%, i.e. the probability of prediction of a type Ia SN being false is 0.7\%. One factor contributing to the high purity is that Ia are the most common type of SNe.

Type Ia supernovae are characterized by the lack of hydrogen emission lines in their spectrum (distinguishing them from type II SNe), as well as the presence of strong silicon lines (distinguishing them from type Ib and Ic SNe). Physically Ia SNe are different from all other classes, as they are believed to be triggered by thermal runaway of a white dwarf reaching the Chandresekhar limit of about 1.4 $M_\odot$ by accreting gas from its giant companion, in contrast to type II, Ib and Ic being triggered by the core collapse of a massive star.

Type Ia SNe are important as cosmological standard candles, because they are a largely homogeneous group, with small variance in their peak luminosity. Nevertheless, there is some variation, but this can be accounted for, thanks to the correlation between the peak luminosity and its rate of decline \citep{phillips1993}. This in turn, once known, can be used to accurately derive the distance to the host galaxy. This has important cosmological applications, enabling e.g. measurements of the accelerated expansion of the Universe, and other cosmological parameters \citep[e.g.][]{riess1998,perlmutter1999}.

\begin{figure}
   \centering
   \includegraphics[width=12cm]{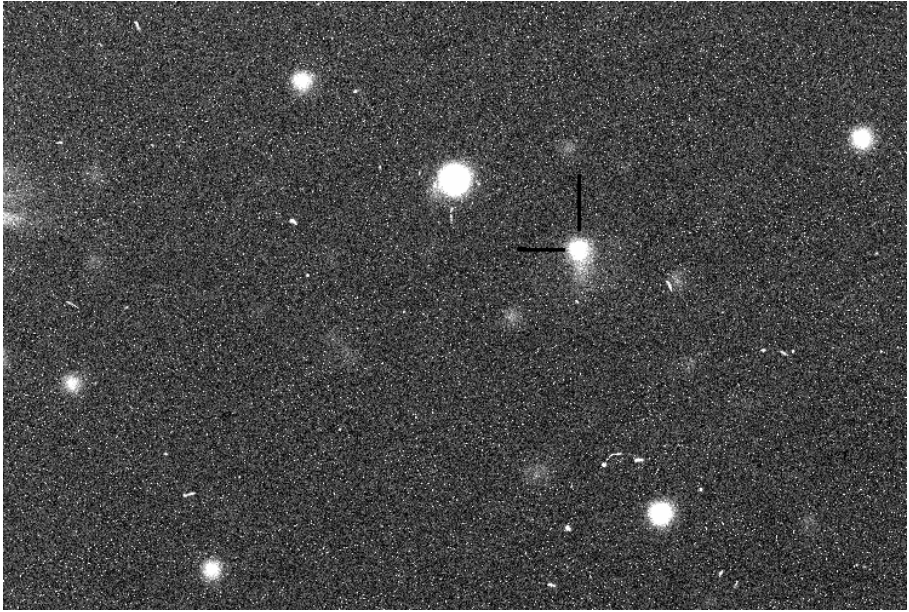}
      \caption{A reduced CCD image of AT 2018we in the R band from Mar 16, composed of two 10 minute exposures. The markings show the SN in the image. Faint light from the host galaxy can be seen around it.}
         \label{sky}
   \end{figure}

A reduced CCD image of AT 2018we is shown in Figure \ref{sky}. We did not find any redshift published for its assumed host galaxy. Thus we have used the peak magnitude of the SN to estimate its distance.

\section{Observations}

We have made differential photometry of AT 2018we with the 0.60 m RC telescope of the University of Helsinki, located in Mets\"{a}hovi, Kirkkonummi, Finland (at coordinates $60^{\circ}13'05''\mathrm{N}, 24^{\circ}23'38''\mathrm{E}$), during Feb-May 2018. As detector we have used SBIG ST-7 CCD. Our photometry is mainly in the Cousins R band, because the object is well visible there, and magnitudes for suitable close by comparison stars have been found in the USNO-A2.0 catalogue \citep{USNO-A2.0}, with some data points also in the Johnson B band, which is affected by our detector being notably less efficient in blue light.

A list of our observations are shown in Table \ref{phot}. A list of comparison stars used is shown in Table \ref{comp}. The chosen comparison stars vary slightly between different nights, depending on the exact field of view.

To correct for galactic extinction and reddening, we subtracted 0.21 mag from the R band observations and 0.35 mag from the B band observations. These values for the coordinates $\alpha$=17h9m57s, $\delta$=82$^\circ$0'3'' are taken from \cite{schlafly2011}. Extragalactic extinction is assumed to be negligible.

\begin{table}
\caption{List of our photometry, corrected for galactic extinction. Exposure times are in R-band if not stated otherwise. Uncertainties are calculated as one standard deviation of the magnitude values given by all the comparison stars used that night. For identifiers of the comparison stars, see Table \protect\ref{comp}.}
\label{phot}
\centering
\begin{tabular}{|c c c c c c|}
\hline
Date & JD-2400000 & Total exposure time & Comparison stars & B & R \\
\hline
Feb 20 & 58170.20 & 25 min & 3, 9, 10, 11 &  & 16.6 $\pm$ 0.2 \\
Feb 21 & 58171.37 & 10 min (R), 10 min (B) & 1, 3, 4, 8, 9, 12, 16 & 16.0 $\pm$ 0.3 & 16.4 $\pm$ 0.2 \\
Feb 24 & 58174.45 & 20 min (R), 20 min (B) & 1, 3, 9 & 15.7 $\pm$ 0.3 & 16.2 $\pm$ 0.1 \\
Mar 16 & 58194.37 & 20 min (R), 20 min (B) & 1, 2, 3, 4 & 17.8 $\pm$ 0.3 & 16.8 $\pm$ 0.1 \\
Mar 27 & 58205.37 & 35 min & 1, 2, 3, 4, 12 &  & 17.0 $\pm$ 0.2 \\ 
Apr 3  & 58212.33 & 32 min & 1, 2, 4, 8 &  & 17.4 $\pm$ 0.1 \\
Apr 13 & 58222.42 & 40 min & 1, 2, 4, 7, 8 &  & 18.0 $\pm$ 0.2 \\
Apr 19 & 58228.35 & 40 min & 1, 2, 3, 13, 14, 15 &  & 18.1 $\pm$ 0.2 \\
Apr 25 & 58234.42 & 55 min & 1, 2, 4, 5, 6 &  & 18.3 $\pm$ 0.1 \\
Apr 27 & 58236.35 & 40 min & 1, 2, 3, 4, 14 &  & 18.3 $\pm$ 0.1 \\
May 5  & 58244.40 & 50 min (R), 85 min (B) & 1, 2, 4, 5, 6, 7, 8 & 19.2 $\pm$ 0.5 & 18.5 $\pm$ 0.1 \\
\hline
\end{tabular}
\end{table}

\begin{table}
\caption{List of comparison stars used for the differential photometry. Magnitudes of the comparison stars are from the USNO-A2.0 catalogue \citep{USNO-A2.0}. The first column shows the identifier we have been using for the star in Table \ref{phot}.}
\label{comp}
\centering 
\begin{tabular}{|c c c c c c|}
\hline 
$\#$ & USNO-A2.0 identifier & RA (J2000) & Dec (J2000) & B-magnitude & R-magnitude \\
\hline 
1 & 1650-01919127 & 17h 9m 45s & $+81^{\circ} 59' 26''$ & 15.9 & 15.1 \\
2 & 1650-01918755 & 17h 9m 29s & $+81^{\circ} 58' 41''$ & 17.8 & 17.4 \\
3 & 1650-01920381 & 17h 10m 37s & $+82^{\circ} 0' 23''$ & 17.4 & 16.3 \\
4 & 1650-01919078 & 17h 9m 43s & $+82^{\circ} 1' 34''$ & 17.0 & 16.8 \\
5 & 1650-01918360 & 17h 9m 14s & $+81^{\circ} 58' 52''$ & 18.1 & 18.0 \\
6 & 1650-01918322 & 17h 9m 12s & $+81^{\circ} 59' 42''$ & 18.0 & 17.3 \\
7 & 1650-01918536 & 17h 9m 21s & $+82^{\circ} 2' 34''$ & 15.0 & 14.0 \\
8 & 1650-01919813 & 17h 10m 14s & $+82^{\circ} 2' 17''$ & 17.4 & 15.6 \\
9 & 1650-01920757 & 17h 10m 52s & $+81^{\circ} 58' 57''$ & 17.4 & 16.8 \\
10 & 1650-01920851 & 17h 10m 56s & $+81^{\circ} 57' 23''$ & 17.7 & 17.1 \\
11 & 1650-01920426 & 17h 10m 39s & $+81^{\circ} 56' 25''$ & 16.5 & 16.1 \\
12 & 1650-01920130 & 17h 10m 27s & $+82^{\circ} 2' 34''$ & 15.4 & 15.0 \\
13 & 1650-01920484 & 17h 10m 42s & $+81^{\circ} 57' 57''$ & 19.5 & 17.3 \\
14 & 1650-01919769 & 17h 10m 12s & $+81^{\circ} 57' 23''$ & 18.8 & 17.6 \\
15 & 1650-01919660 & 17h 10m 7s & $+81^{\circ} 56' 59''$ & 18.7 & 17.2 \\
16 & 1650-01921356 & 17h 11m 17s & $+82^{\circ} 0' 41''$ & 16.4 & 16.0 \\
\hline 
\end{tabular}
\end{table}

\section{Data analysis}

We have used IRAF (Image Reduction and Analysis Facility) for the reduction and analysis of our data. Basic bias and flat field corrections were made. We then stacked together 2 to 9 individual exposures for each night with the task $imcombine$, to gain the final image which we used in the photometry.

Then, the final magnitude has been derived as a mean of the values given by using different comparison stars visible in the image that particular night. An uncertainty of a few tenths of a magnitude is caused by inconsistencies between the comparison stars, since their status as 'standard stars' is not confirmed in any way. Nevertheless, we have gained reasonable results with acceptable error margins for our purposes.

Since the host galaxy significantly contributes to the total light, aperture photometry is not reliable for the SN. We thus made a PSF (Point Spread Function) fit for isolated field stars, which was then applied to the SN. Here, we used the $daophot$ package of IRAF. We then calculated the magnitude as the mean of magnitudes given to the SN by different comparison stars, and the uncertainty as one standard deviation of the results. The results of our photometry is shown in Table \ref{phot} and Figure \ref{LC}.

To the R band light curve, with magnitude $m$ at time $t$, we fitted an empirical function of the form

\begin{equation} \label{func}
m=\frac{f_0 + \gamma(t-t_0) + g_0 \exp(\frac{-(t-t_0)^2}{2\sigma_0^2}) + g_1 \exp(\frac{-(t-t_1)^2}{2\sigma_1^2})}{1-\exp(\frac{\tau-t}{\theta})},
\end{equation}

\noindent as discussed by e.g. \cite{vacca1996} and \cite{contardo2}.

This fits four different functions to the light curve: first, an exponential to the rising phase before light maximum, which approaches unity near the maximum. This exponential is multiplied with the sum of two Gaussians, first one for the maximum, and a second one to the bump in the ligt curve seen around 30 days after maximum, and a linearily decreasing line which describes the later parts of the fading light curve. Here $t_0$ is the phase where the decreasing line and the first Gaussian are normalized, and the second Gaussian is normalized at $t_1$. $g_0$ and $g_1$ are the amplitudes of the two Gaussians, and $\sigma_0$ and $\sigma_1$ their widths. $f_0$ is the intercept of the line, $\gamma$ the slope and $\theta$ and $\tau$ the characteristic time and phase zero point of the exponential. Best values for these parameters were fitted.

There are 10 free parameters in the fit, which is of the same order as the amount of data points in our light curve, so one should not assume the model to precisely describe reality, since by adjusting the free parameters a multitude of different fits could be done, all fitting the sparse data sufficiently well. Therefore, one must check that the fit seems physically reasonable, since many combinations of the free parameters can yield good fits, which are physically unreasonable. Sensible initial guesses and parameter restrictions for the fitting routine are discussed by \cite{contardo2}.


\section{Results}

\subsection{R band light curve}

   \begin{figure}
   \centering
   \includegraphics[width=12cm]{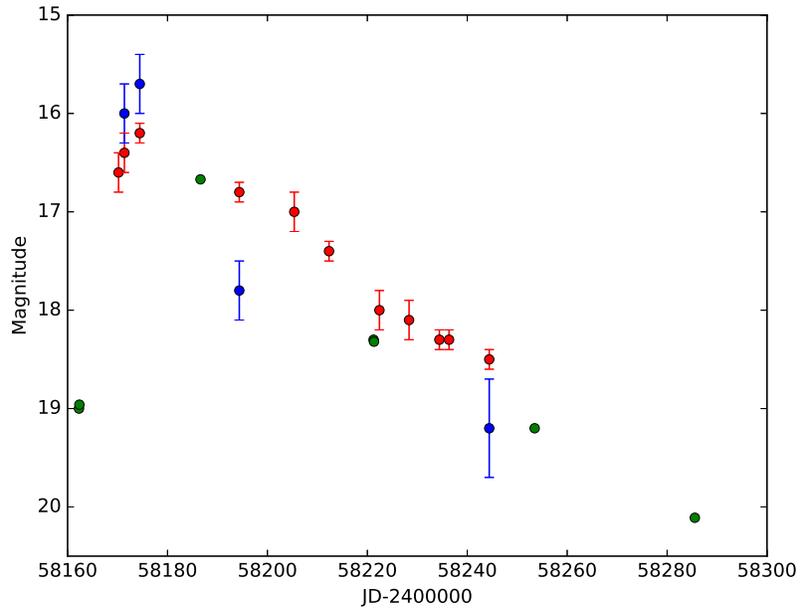}
      \caption{Light curve of the photometric observations, corrected for galactic extinction. Red dots are in the R band, and blue in the B band. For comparison, the green dots are Gaia data in the Gaia G broad band white light filter, available at the alert web page (http://gsaweb.ast.cam.ac.uk/alerts/alert/Gaia18alj/).}
         \label{LC}
   \end{figure}

   \begin{figure}
   \centering
   \includegraphics[width=12cm]{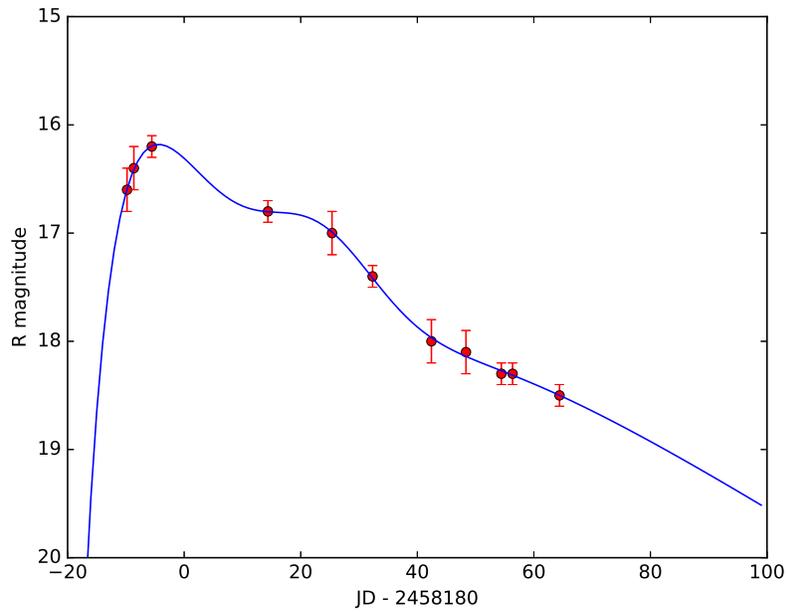}
   \caption{A function of the form of Equation \ref{func} fitted to our R band data.}
         \label{fit}
   \end{figure}

A light curve constructed from our photometry is shown in Figure \ref{LC}. The initial prediction for AT 2018we was a type Ia supernova. The light curve resembles that of a typical SN of type Ia. Some characteristic features seen in R band light curves of Ia SNe are the fast decline after the peak luminosity, after which a faint second peak is seen at around 30 days after maximum, caused by cobolt isotopes with slower decay times than the nickel isotopes powering the maximum. Although the signal of the secondary peak is weaker than our error margins, it fits very well into expected light curves for Ia SNe, as described e.g. by \cite{contardo1}. The secondary peak is also a weak suggestion that AT 2018we is indeed of type Ia and not Ib or Ic, whose light curves are similar, but lacking this feature in the R band (see e.g. \cite{drout,taddia2018}). After this the luminosity continues to decrease, but now the slope is flatter. In the B band, on the other hand, the luminosity decline is much faster than in the R band.

We fitted a function of the form of Equation \ref{func} to the R band light curve. The resulting fit is shown in Figure \ref{fit}. The fitted values for the parameters in Equation \ref{func} are shown in Table \ref{param}.

\begin{table}
\caption{The best fit values for the free parameters in Equation \ref{func} for the R band light curve.}
\label{param} 
\centering 
\begin{tabular}{|c c|} 
\hline
Parameter & Value \\
\hline 
$f_0$ & 15.295 mag \\
$\gamma$ & 0.034 mag/d \\
$g_0$ & -8.0 mag \\
$t_0$ & -24.803 JD-2458180 \\
$\sigma_0$ & 21.046 mag \\
$g_1$ & -0.787 mag \\
$t_1$ & 23.201 JD-2458180 \\
$\sigma_1$ & 9.636 mag \\
$\tau$ & -27.068 JD-2458180 \\
$\theta$ & 20.0 d \\
\hline 
\end{tabular}
\end{table}

\subsection{B band observations as a distance indicator}

Our observations in the B band are quite sparse and similar light curve parameters cannot be derived from them as for the R band, but they can be used to estimate the distance to the SN and the host galaxy from the peak luminosity, which is most commonly done in the B band.

Even if type Ia SNe are largely homogeneous, there are variations in their peak luminosities, probably due to differences in their compositions. Indeed, they are not perfect standard candles, but rather 'standardizable candles'. There is a correlation between the maximum absolute luminosity and the rate of decline of the light curve, called the Phillips relation. This relation is named after \cite{phillips1993}, who derived the following form for it in the B band:

\begin{equation}
M_{\mathrm{B,max}} = -21.726+2.698 \Delta m_{\mathrm{15}}(B).
\end{equation}

\noindent Here, $\Delta m_{\mathrm{15}}(B)$, introduced by \cite{phillips1993}, is the amount of magnitudes the SN fades in 15 days after the maximum in the B band.

At Feb 24 the B magnitude was 15.7, which, when estimating the light curve by eye, seems to be close to maximum. The separation to the next data point is 19.92 days, so linearly interpolating the value of $\Delta m_{15} (B)$ may not be too far stretched. This would yield $\Delta m_{15}(B) \approx$ 1.58, which would make the SN a fairly quickly declining one, and thus probably rather a faint one than a bright one, in absolute terms. The value for $\Delta m_{15} (B)$ should be regarded as a minimum value, since the precise timing of the peak magnitude in the B band cannot be deduced accurately from our sparse data, but it most likely ocurred after Feb 24. Also, the timing of the peaks in the B and R bands might not be exactly the same.

By using $\Delta m_{\mathrm{15}} (B) = 1.58$, we get $M_{\mathrm{B,max}}$ = -17.46. Now we can estimate the distance $d$ to the host galaxy by solving the distance modulus:

\begin{equation} \label{dist}
d = 10^{\frac{m_{\mathrm{B, max}} - M_{\mathrm{B, max}}}{5}} \cdot 10 \mathrm{pc},
\end{equation}

\noindent where $m_{\mathrm{B, max}}$ is the apparent magnitude at peak luminosity, corrected for galactic extinction.

Our observations in the B band are too sparse to give an accurate result, but we can estimate the value 15.7 from Feb 24 to be close to maximum. Thus, by inserting $m_{\mathrm{B, max}} = 15.7$ to Equation \ref{dist} one would get the distance to the SN as $d \approx 43$ Mpc, assuming the SN follows the Phillips relation. This value should anyway be regarded only as a guide line, so we did not estimate any error bars.

A further source of error, not considered here, might be extinction and reddening caused by dust in the host galaxy.

Our relevant results about the SN derived from the B observations are summarized in Table \ref{summ}.

\begin{table}
\caption{Our results for the relevant parameters of AT 2018we derived from the B data.}
\label{summ}
\centering
\begin{tabular}{|c c|}
\hline 
Parameter & Value \\ 
\hline 
$\Delta m_{\mathrm{15}} (B)$ & 1.58 \\
$M_{\mathrm{B, max}}$ & -17.46 \\
Distance & 43 Mpc \\
\hline
\end{tabular}
\end{table}


\section{Conclusions}

Based on our observations, the initial classification of AT 2018we as a type Ia supernova is supported. Although there is no spectrum of the SN, which would be a more definitive proof of its type, the light curve resembles that of a typical type Ia SN. In contrast to type II SNe, however, the light curves of type Ib and Ic SNe resemble those of type Ia, so it might not be completely safe to rule out these types of core collapse SNe for AT 2018we. The weak secondary bump around 30 days after maximum, seen in the R band light curve, would still suggest that the initial GS-TEC prediction was correct, and AT 2018we is of type Ia.

We have also estimated the distance to its assumed host galaxy as 43 Mpc, assuming AT 2018we to be of type Ia. The host is thus apparently a relatively closeby galaxy.

\section*{Acknowledgments}

We acknowledge ESA Gaia, DPAC and the Photometric Science Alerts Team (http://gsaweb.ast.cam.ac.uk/\\alerts). We are grateful to Akke Viitanen for his helpful comments, which considerably helped to improve the paper.

\bibliographystyle{plainnat}
\bibliography{2018we}

\end{document}